*Invited Paper*

# New Frontiers of Network Security: The Threat Within

Sugata Sanyal[1], Ajit Shelat[2], Amit Gupta[2]

*1: Tata Institute of Fundamental Research, Mumbai India*
sanyal@tifr.res.in

`2: Nevis Networks India Pvt. Limited`
ajit.shelat@nevisnetworks.com
amit.gupta@nevisnetworks.com

*Abstract*— Nearly 70% of information security threats originate from inside an organization. Opportunities for insider threats have been increasing at an alarming rate with the latest trends of mobility (portable devices like Laptop, smart phones and iPads etc), ubiquitous connectivity (wireless or through 3G connectivity) and this trend increases as more and more web-based applications are made available over the Internet. Insider threats are generally caused by current or ex-employees, contractors or partners, who have authorized access to the organization's network and servers. Theft of confidential information is often for either material gain or for wilful damage. Easy availability of hacking tools on the Internet, USB devices and wireless connectivity provide for easy break-ins. The net result is losses worth millions of dollars in terms of IP theft, leakage of customer / individual information, etc. This paper presents an understanding of Insider threats, attackers and their motives and suggests mitigation techniques at the organization level.

## I. INTRODUCTION

It is a well established fact that 70%+ of threats to an organization's network and network-based infrastructure originate from inside. US CERT in their recent study has predicted that malicious internal users deploy root kits, carry out identity thefts, use spyware and gain unauthorized access. Cyber attacks with financial motives are based on a mix of strategies and mechanisms ranging from social engineering to viruses and viruses have increased dramatically in the recent past. Till date, to a very large extent, the organizational security focus has generally been on protection of the perimeter and the end points. The LAN itself has lacked sufficient security provisions. With an increasing laptop population, organization-wide wireless connectivity and increasing number of remote sites to the organizational network, the traditional approach of fortifying the perimeter has diminished in providing adequate security. Now, there is a need to protect each and every network object rather than just perimeter or PCs. Newer technologies like Network Access Control (NAC), Data leakage prevention (DLP) and organization wide threat monitoring and alerting system along with policy enforcement are required to mitigate any threat or block any suspicious activity within the organization. Insider threat mitigation solutions are required for the following business challenges:

  1. Preventing unauthorized system access to critical IT resources
  2. Preventing data breaches
  3. Preventing sabotage (internal)
  4. Preventing theft of intellectual property
  5. Reduction of administration cost related to security
  6. Better audit ability to address compliance requirements

## II. WHAT HAS CHANGED

In the last few years a few trends have been clearly visible. Portable devices, i.e. laptops, net books and smart phones, have become popular due to lower prices, better form factor, ease of use, etc. They also offer the advantage of mobility in terms of 'anytime, anywhere computing'. A certain class of employees need these kind of devices to carry out their work effectively and efficiently - sales or technical support personnel can work from anywhere by connecting over the Internet, typically using VPNs, to connect to the corporate networks.  These mobile devices not only have sensitive organizational data stored on them but they also provide easy access to corporate networks – hence, they are a potential source of data loss  - deliberately or otherwise.

Similarly, employees located inside an organization and connected to the corporate network can send out critical and confidential information using the company's Internet facility. This can be done using peer-to-peer applications of various types. Some of these applications can be of very high security risk due to their ability to operate undetected within the network. Applications/utilities such as Bittorent are very popular for downloading large volumes of data (such as movies and music). Such large volumes of data moving over a LAN can choke the LAN infrastructure.  Such applications use many techniques to evade detection – such use of dynamically assigned ports as well as using popular/well known TCP ports (such as SMTP, ftp etc.) which are not normally blocked by an organization. Normal firewalls or security equipment cannot detect these applications. Cisco has predicted that by 2014 Video will contribute to 91% of data volumes over the Internet.

The various types of USB based removable storage devices now provide extremely large capacity storage. Being small in size, the company's confidential data can be easily taken out on a single drive without anyone noticing it.

The Internet provides access to thousands of free and low cost hacking tools. These tools can be easily downloaded and tried by just about any computer literate person without deep knowledge of programming etc. These tools can break passwords of desktops, databases, servers, switches and routers on either local or distant networks. These tools can help deploy key loggers or obtain screen shots at regular intervals from the desktop where deployment has been done remotely.

With the popularity of social networking, employees are spending hours and hours on sites like Facebook, using Internet connections to play games, engaging in stock-market trading and downloading contents, not in line of their business requirements. This costs organizations millions of dollars in terms of lost business and resources.

Wireless offers convenience and lower cost connectivity, it is, however, not as secure as wired networks. Due to its advantages, wireless connectivity is growing rapidly. This also makes corporate wireless networks vulnerable. A recent example, showed techniques which can break wireless WPA-II security (presented in the last Black Hat conference).

### III. WHAT IS AN INSIDER THREAT?

Insider Threat is about breaching the confidentiality, integrity and availability of organizational information system using authorized or unauthorized access. These attacks are carried out by current or ex-employees, business partners or contractors. The net result of insider attack amounts to losses to the organization.

### IV. MOTIVATION FOR INSIDER THREATS

Insiders have knowledge and access to internal information. As most of organizational IP information is, nowadays, resident in digital form and is stored across different servers, this potentially valuable information is easily accessible to employees to access and then leak to competitors, etc. As a typical example, in 2007, Gary Min was sentenced to jail for stealing approximately 16,000 documents and 22,000 scientific abstracts from his employer, DuPont. Gary had collected these documents from various servers using his authorized access and gained unauthorized access using tools. One of the most common motivations for carrying out an insider attack remains financial gain.

Sometimes, disgruntled employees causing harm or damage are also seen as a reason. On 9$^{th}$ Aug 2010, Terry Childs received a sentence of 4 years of imprisonment. Childs, effectively held San Francisco's Fibre Wide Area Network (Fibre-WAN) hostage after giving himself sole access to the administrative passwords, which he refused to hand over to the city when requested by his supervisor and caused 900,000 US$ of losses during a 12 day standoff.

As Cyber warfare is becoming an integral part of military strategies across the world, nations are also deploying large quantities of resources to infiltrate adversary networks and infrastructure with the aim of either stealing valuable information or holding the adversary's infrastructure hostage. Often, custom spyware ('bots') is used to infect a system, which further infects the entire network. And these bots can then keep sending critical internal information to the command centre and/or harm or obstruct the operation of the infected system.

### V. TECHNIQUES USED FOR INSIDER ATTACKS

An insider with malicious intent can use a variety of the following techniques.
a. Social Engineering – One can obtain a colleague's password by looking over the shoulders or by enabling remote management to gain access, when he/she has left their computer unattended. The attacker can also insert malicious software to extract password etc. This password is later used for break-ins. Common user accounts which are used for testing, training etc, can also be used for break-ins.
b. Tools Usage – A large variety of hacking tools are available i.e. root kits, password crackers, password sniffers and remote key logger installers.
c. Misuse of Privilege – an administrator can deliberately provision a freshly installed computer with privileged access enabled – to be used at a later time by him/her. Alternately, backdoor accounts can also be created.
d. Installing Modem - Data cards can be inserted into a system so as to provide Internet access to enable sending information outside the organization. Other mechanisms generally employed use private mails or encrypted communication to the external world.
e. Wilful Deletion of Information – authorized users remove or modify databases and commit fraud at a later date.
f. Installing Rouge Devices – i.e. installing a rogue wireless access points so that unauthorized network connections can be made. Or leaving a rogue laptop on the network to sniff network activity to be used for later break-ins.
g. Using Denial of Service Attacks – internal resources can be swamped with unwanted traffic to bring the services down.
h. Breaking into Wireless Network, carrying out man –in –the-middle attacks or using sophisticated attacks using impersonation i.e. User/ System / Switch / VLAN using tools to gain authorized access.
i. Often, useful information is hosted on internal web servers and these servers are broken in with variety of internal attacks.

With easily available tools, the above attacks can be carried out with ease and these need only elementary computer and network knowledge.

VI. SOLUTIONS FOR MITIGATION OF THESE ATTACKS

Most attack tools and malware exploit the vulnerabilities that exist in Operating Systems, software applications and on the networks. Other type of attacks exploits poor security policies within an organization. Having good security policies and poor implementation also leads to information losses through insider attacks. Organizations are subject to such vulnerabilities due to inadequate security reviews and monitoring and absence of an alerting mechanism within an organization. To ensure organizations pay necessary attention to the information security, various legal compliances are required to be met – for e.g. SOX, HIPAA, PCI, ISO27000, etc. These are various compliance standards to be met: Surbanes-Oxley (SOX) is for business compliance, HIPAA is for medical records and information, PCI is for credit card and related transactions/sales (used in retail chains, etc.), ISO27000 is a corporate security requirement standard. Most Governments are taking steps to ensure that Information Security standards are maintained and is making legally bound for businesses. Information Security System certification is also becoming popular, by which organizations can confirm to their business partners that they have effective information security systems in place.

Having Information Security policy, well understood across the entire organization is the first step towards security. Effective enforcement of this policy is the next important step with regular monitoring and alerting mechanisms.

Today's threats have become increasingly sophisticated. These new threats have driven the need for a layered approach to security that employs multiple threat detection and prevention techniques before, during and after any user attempts access to the networks. This includes endpoint integrity verification, network authentication, encryption and access control, application-layer firewall, signature-based intrusion prevention, anomaly-based threat detection, and others.

As there is no silver bullet for information security, organizations have to rely on multilayered security solutions from different vendors. These solutions should appropriately cater to the organizational security needs but should address security from end to end. Often, security planning takes a back seat as security is considered not having an ROI (Return on Investment). But that perception is changing very fast. Companies have seen improvement in their procurement prices and better sales realization with improved organization wide security. The lack of security can result into penalties, besides loss of brand in the market.

Protecting the organizational information assets from insider threats requires both proactive and reactive security measures.

Proactive measures are applied before an endpoint or user is granted access to the network, after an appropriate authentication and after its security posture checks. These checks ensure that the user system is a corporate resource and it is not trying to connect its personal system to the corporate network. To overcome the problem of passwords, techniques like dual factor authentication are used. Dual factor authentication use dynamic passwords, which keep changing with time. At many places biometric authentication is also being used to ensure passwords are not misused and to ensure that a person can be held responsible. When password management is in place, then single sign-on solutions ensure that the user gets access as per policy and as per his/her role. The log management provides usage logs and alerts for any abnormal behavior.

One needs to ensure that the Anti-Virus signatures, data files and security patches are up to date. Users on the network are automatically patched up and updated with anti virus signatures using central patch management solutions and Anti Virus systems.

To check and enforce the security policies, a concept termed Network Access Control (NAC) is becoming essential. The job of checking End point security posture, integrating with authentication, patch and Anti virus management is done by NAC. Users are allowed access to network resources only if the desired security policies have been adhered to. NAC works at the network layer and continuously checks the security posture.

After connection to the network, security also entails Intrusion Prevention System to be deployed in the LAN to block any malicious (malware or hacking) activity. Malicious User can automatically be 'thrown out' of the network or be quarantined.

To ensure that data is not leaked out from the endpoints, most end point security solutions provide integrated feature list that includes Anti-Virus, Anti-Spyware, Host based Intrusion Prevention Systems (HIPS), Data Leakage Prevention (DLP) software and personal firewall. Data leakage prevention provides control on removable USB drives as per policy as well as measures to prevent theft of laptops, etc.

A number of coordinated and infected PCs can launch DoS (denial of service attacks) or DDoS (distributed denial of service attacks) attacks. These attacks are designed to slow down or disable networks altogether. These attacks are among the most serious threats that organization networks face. These attacks need to be neutralized in real-time.

The crucial requirement to meet the insider threat is to provide very high performance and deep security. Most of the networks today work on Gigabit Ethernet with current desktops/Laptop come equipped with Gigabit Ethernet port with very high computing power.

Besides having above multi layered solution in place, there is also a need to have security correlation between different security applications as a part of monitoring and review. This correlation should be able to generate real-time alerts to take preventive action.

## VII. IMPACTS OF FUTURE TRENDS ON SECURITY

Devices like iPads and Smart phones with 3G capabilities are becoming an accepted part of corporate IT infrastructure as a converged IT device for both voice and data traffic. These are likely to connect to corporate network and servers from public networks like Internet using VPN. The Corporate Information Security group will have the responsibility of protecting these mobile users as well as corresponding corporate information. Being on the public network, these users and devices will need protection from increasingly potent malware and hackers.

Higher speeds on wired or wireless network with higher speed processors will result in richer applications. These new applications will pose newer security risks.

Running these rich applications on high speed network will require security devices that have higher performance along with deep packet inspection capabilities.

## VIII. CONCLUSIONS

With changes in technology landscape the insider threats are rising at an alarming rate. Most of insider threats are far more sophisticated and are difficult to catch. As attack is carried out with inside knowledge, it is generally very expensive for the organization to catch these. Often, insider attacks are not even caught as attacker leaves no traces. To stop these attacks both proactive and reactive approach is required. The proactive approach implies automatic enforcement of security policies across the organization along with an effective monitoring, forensic and logging mechanism. Each element of the network has to be protected rather than the usual approach of concentrating on perimeter security. Multilayered security with event correlation is recommended to stop any attack on the resources. Real-time monitoring and alerting in case of any malicious behavior, forms the part of reactive strategy.